\UseRawInputEncoding
\documentclass{article}
\usepackage{graphicx}
\usepackage{textcomp}
\usepackage{amssymb}
\usepackage{amsmath}
\usepackage{bm}
\usepackage{mathtools}
\usepackage{nicefrac}
\usepackage{subfig} \usepackage{xcolor}
\usepackage{tikz}
\usepackage{pgfplots}
\usepackage{ulem}
\usepackage{cite}

\newcommand{\nQR}{n^{(\ell\lambda)}_{QR}}
\newcommand{\tQR}{t^{(\ell\lambda)}}

\newcommand{\omg}{\omega_{xt}^{(\ell\lambda)}}
\newcommand{\trace}{\lambda_{d,xt}}
\newcommand{\epsxt}{\epsilon_{xt}}
\newcommand{\RR}{{(R)}}
\newcommand{\I}{{(I)}}
\newcommand{\II}{{(II)}}
\newcommand\redsout{\bgroup\markoverwith{\textcolor{red}{\rule[0.5ex]{2pt}{1pt}}}\ULon}
\newcommand{\tred}{\textcolor{black}}
\newcommand{\tblue}{\textcolor{black}}
\newcommand{\tgreen}{\textcolor{black}}
\pgfplotsset{compat=1.14}

\begin{document}


\title{A Local-Realistic Quantum Mechanical Model of Spin and Spin Entanglement}

\author{Antonio Sciarretta}
\date{}

\maketitle

\begin{abstract}

This paper aims at reproducing quantum mechanical (QM) spin and spin entanglement results using a realist, stochastic, and local approach, without the standard QM mathematical formulation.
The concrete model proposed includes the description of Stern-Gerlach apparatuses and of Bell test experiments. 
Single particle trajectories are explicitly evaluated as a function of a few stochastic variables that they assumedly carry on. 
QM predictions re retrieved as probability distributions of similarly-prepared ensembles of particles.
Notably, it is shown that the proposed model, despite being both local and realist, is able to violate the Bell--CHSH inequalities by exploiting the coincidence loophole and thus intrinsically renouncing to one of the Bell's assumptions.
\end{abstract}


\section{Introduction}
\label{intro}

Efforts to provide a fundamentally realist and causal description underlying the abstract formalism and reproducing the inherently stochastic predictions of quantum mechanics have been attempted since the early days of quantum mechanics itself \cite{santos, harrigan, einstein, despagnat, budiyono}.
However, Bell's theorem and its descendants \cite{bell, clauser} are regularly used to dismiss any possibility that a local realist quantum mechanical model could even exist.

Despite its mathematical simplicity, interpretation of Bell's theorem has given rise to a vast literature, in particular concerning its assumptions and the conclusions that can be drawn.

The usual assumptions used in deriving Bell inequalities are realism (properties of physical systems are elements of reality, outcomes of tests are determined by some hidden variables), factorability (these outcomes cannot be influenced faster than the speed of light), and measurement independence (the measurement setting choices are independent of the hidden variables and vice versa) \cite{jarrett, shimony, larsson, vervoort, budiyono}. 

However, all experimental demonstrations that attempt to violate Bell's inequality \cite{aspect, weihs} have to deal with practical problems ('loopholes') and therefore require additional assumptions in order to reject local realism.
In principle, any violation could be caused by the failure of these additional assumptions, rather than by local realism itself \cite{larsson}.
Consequently, several experiments have been conducted with the purpose of observing violations of Bell's inequalities that are as much as possible loophole-free \cite{aguero,christensen,giustina,shalm}.

In summary, after more than fifty years from Bell's original paper, there is no real consensus on several interpretational issues. 
In particular, dismissing realism and locality all short as a conclusion of BT remains unjustified to many researchers.
%
Notwithstanding, since Bell inequalities are experimentally violated, at least one of the Bell's assumptions above must be false.
Rejection of one particular of these assumptions corresponds to one of the admissible interpretations or solutions of Bell’s theorem.

The standard approach ('indeterminism') is to reject 'realism', that is, the existence of any hidden variable (HV) completing quantum mechanics and thus the fact that the values of the outcomes 
even exist before their measurement.
Another possible solution is to reject factorability. Since this assumption is often, probably incorrectly \cite{budiyono}, equated to no-signaling and thus locality \cite{vervoort}, such approach leads to non-local theories that have many advocates (e.g., Bohmian mechanics).
The last possibility concerns the validity of measurement independence (MI). It is often believed that MI represents the freedom of the experimenter to choose the measurement setting at will and thus is also referred to as freewill hypothesis. The fact that MI is not satisfied have been often explained by some kind of (super)determinism or ``conspiracy". 
Altogether, other, less unpleasant reasons to renounce to these assumptions exist.

This promising approach consists to ``exploit the loopholes" of Bell's theorem. In other words, a model can be derived that explicitly takes into account those supposed imperfections of experimental procedures that, instead of being desirably eliminated, constitute a fundamental prerequisite of the observed correlations and help recovering the quantum statistics \cite{pearle,fine,pascazio,brans,khrennikov,deraedt2016}.

For example, one of these proposals explicitly uses the ``detection" loophole by assuming that the probability of joint detection (detector efficiency) depends on the settings \cite{thompson,selleri}. 
The ``contextuality" loophole \cite{nieuwenhuizen,khrennikov08}, for which hidden variables that supposedly affect the detectors would be differently distributed for different settings, is another possibility that, however, has not been embodied in a concrete model to my best knowledge \cite{vervoort}.
%
Similarly, although non-ergodic \cite{buonomanoEPR,khrennikov2} solutions would in principle belong to this category, no proposal has been issued in this direction as per the best knowledge of the author. 
Other \textit{ad hoc} attempts include mathematical artefacts that correctly reproduce the QM correlations in an abstract and physically unexplained fashion \cite{mueckenheim,dilorenzo}.

%
However, the most interesting attempt to reproduce Malus' law and QM predictions in a local-realistic context is the event-based class of models exploiting the ``coincidence" loophole proposed by \cite{scalera,larsson,michielsen,deraedt,zhao,deraedt2016}.
In this approach, the key role is played by the time delay between particle arrivals at the detectors of a Bell-type experiment, so that coincidences are counted only if two particles arrive at roughly the same time.
In a recent development \cite{deraedt2017}, this approach is extended to the ``photon identification" loophole.
Aimed at providing a counterexample to usual solutions of BT, the approach of \cite{scalera,larsson,michielsen,deraedt,zhao,deraedt2016,deraedt2017} is not intended to represent a fundamental sub-quantum mechanism. Time delays are heuristically designed or justified by invoking properties of the measurement apparatus.
%

This paper explores the possibility of providing a local-realist sub-quantum mechanism that copes with BT by pushing the aforementioned approaches to a more fundamental level via an intangible loophole.
Inspired by first principles, spin/polarization properties are accommodated and integrated with momentum-related ones in order to recover other typically quantum behaviors.

%
In \cite{2017,compa}, I have already proposed a model mimicking quantum mechanics (QM) of \textit{spinless} particles with local, realist, and stochastic features.
%
The \textit{stochastic} behavior that is manifested by the empirical evidence of QM is explained by assuming a fundamental randomness both in preparation and in particles trajectories. The emergence of QM behavior is a consequence of the particular rules of motion chosen. The motion of individual particles and their interaction with external forces take place on a discrete space–time under the form of a lattice. Particle trajectories are asymmetric random walks, with transition probabilities being simple functions of a few quantities (playing the role of hidden variables) that are either randomly attributed to the particles during their preparation, or stored in the lattice nodes that the particle visits during the walk. 
%
The lattice-stored information is progressively built as the nodes are visited by successive emissions. This process, where particles leave a ``footprint" in the lattice that is used by subsequent particles implies that the interactions between subsequent emissions in an ensemble fulfill \textit{localism}, albeit through the mediation of the lattice. Quantum behavior emerges for an ensemble of similarly-prepared particles as a consequence of this mechanism.

The main characteristics that distinguishes quantum spin from classical magnetic moment behavior is probably the quantization of the former after a measurement, e.g., by a Stern-Gerlach (SG) apparatus, is performed. 
This behavior is described in standard QM using matrices and eigenvectors. 
In alternative theories, spin has been derived from path integrals \cite{nielsen} and stochastic mechanics \cite{dankel,garbaczewski}. %
The local-realistic mechanism proposed here for spin involves a few additional hidden variables that are assumed to be carried on by particles of the ensemble.
These variables are subject to stochastic preparation at sources and time evolution, including interaction with the lattice nodes storing the information about the magnetic field.
When considering ensemble probabilities, the fundamental mechanism leads to recover Malus' law and the standard correlation statistics of the singlet state, including violation of Bell's inequalities.

Coincidences at detectors are counted based on arrival time. This model feature is in line with previously published material that showed violations of BI for pairs of momentum-entangled particles whose arrivals are counted at particular spatiotemporal lattice nodes representing the detectors \cite{2017,compa}.
The model proposed does not use \textit{ad hoc} expressions for arrival times, which instead naturally follow from the momentum carried on by particles. Indeed, time coincidences are retrieved as a consequence of a more general energy equivalence.

The paper is organized as follows. 
In Sect.~\ref{sec:spin} the model rules for spin $1/2$ are introduced, both for homogeneous and inhomogeneous (SG) fields. Two-particle spin entanglement is discussed in Sect.~\ref{sec:entang}. 
\tgreen{Two} appendices complement the paper. In Appendix~\ref{sec:summ} a summary of the spinless model is presented. One-half spin results are extended to higher spins in Appendix~\ref{sec:higher}. 

\section{Spin 1/2 Model} \label{sec:spin}
We describe an ensemble of particles that are emitted at a source after having been similarly prepared.
%
Each emission evolves on the nodes of a discrete spatiotemporal lattice. The lattice is composed of three spatial dimensions $x=\{x_d\}\in\mathbb{Z}^3$, and one temporal dimension $n\in \mathbb{N}$. Each of the dimensions is characterized by a fundamental length (the spatial dimensions share the same value) and acts independently.

\subsection{Microscopic model} \label{sec:spinmot}

Particles are emitted at source node ($x_0$ and $n_0$) with randomly-attributed properties denoted as ``source spin", $s_0\in[-1,1]$ and ``source polarization", $\mu_0=\{\mu_{0d}\}\in \mathbb{Q}^3$, such that $\sum_{d=1}^3 \mu_{0d}^2 =1$.
While $s_0$ remains constant during a particle's evolution, polarization is prone to change at each time the particle experiences a magnetic field, which is represented under the form $B=\{\beta_d \} B_M$, such that $\sum_{d=1}^3 \beta_d^2=1$. Clearly, $\beta$ represents the unit vector along which the physical field is directed. The quantity $B_M$ represents the magnitude of the magnetic field in lattice units.

The evolution of the polarization follows the rule
%
\begin{equation}
\mu_d[n+1]=\mu_d [n]-\gamma\mu_M B_M \left(\mu_{d+1} [n] \beta_{d+2} [n]-\mu_{d+2} [n] \beta_{d+1} [n] \right)\;,
\label{eqn:polarev}
\end{equation}
with $\mu_d [n_0]=\mu_{0d}$. The quantity $\mu_M$ represents the magnitude of the magnetic moment of the particle, and the dimension indexes must be taken as modulo three. The quantity $\gamma$ represents the gyromagnetic ratio.
\tred{Overall, (\ref{eqn:polarev}) mimics the QM equation for spin precession.}
Note that, if the $\beta_d$'s are constant, the sum $\sum_d\mu_d^2$ is constant, too.
Similarly, the \tred{scalar product}
\begin{equation}
 M[n]:=\sum_d\mu_d[n]\beta_d[n]  \in[-1,1]
\label{eqn:momprop}    
\end{equation}
does not change \tred{during the evolution} if the field is constant  \tred{and naturally} corresponds to the cosine of the angle between the two directions $\mu_0$ and $\beta$.

\begin{figure}
\centering
\subfloat []{
\begin{tikzpicture} [scale=0.65]

\draw [->] (3,-0.5) -- (3,6.5); 
\node [above] at (3,6.5) {$s$};
\draw [->] (-0.5,3) -- (6.5,3); 
\node [right] at (6.5,3) {$s_0$};
\draw [gray] (0,0) -- (0,6) -- (6,6) -- (6,0) -- (0,0);
\draw [gray] (0,0) -- (6,6);
\node [below, fill=white] at (6,0) {$1$};
\node [below, fill=white] at (3,0) {$0$};
\node [below, fill=white] at (0,0) {$-1$};
\node [below, fill=white] at (2,0) {$-M$}; 
\node [left, fill=white] at (0,6) {$1$};
\node [left, fill=white] at (0,3) {$0$};
\node [left, fill=white] at (0,0) {$-1$};
\draw [blue] (0,1) -- (5,6);
\draw [<->] (1,2) -- (1,1);
\node [right] at (1,1.5) {$M$}; 
\draw (0,0) -- (2,0) -- (2,6) -- (6,6);
\end{tikzpicture}
}
\subfloat []{
\begin{tikzpicture} [scale=0.65]
\fill [lightgray] (1.5,0) rectangle (2,6);
\draw [->] (3,-0.5) -- (3,6.5); \node [above] at (3,6.5) {$s$};
\draw [->] (-0.5,3) -- (6.5,3); \node [right] at (6.5,3) {$s_0$};
\draw [gray] (0,0) -- (0,6) -- (6,6) -- (6,0) -- (0,0);
\draw [gray] (0,0) -- (6,6);
\draw (0,0) -- (2,0) -- (2,6) -- (6,6);
\draw (1.5,0) -- (1.5,6) -- (2,6);
\node [below, fill=white] at (6,0) {$1$};
\node [below, fill=white] at (3,0) {$0$};
\node [below, fill=white] at (0,0) {$-1$};
\node [below left, fill=white] at (1.5,0) {$-M'$}; 
\node [below, fill=white] at (2,0) {$-M$}; 
\node [left, fill=white] at (0,6) {$1$};
\node [left, fill=white] at (0,3) {$0$};
\node [left, fill=white] at (0,0) {$-1$};
\draw [blue] (0,1) -- (5,6);
\draw [red] (0,1.5) -- (4.5,6);
\draw [<->] (4,5) -- (4,5.5); 
\node [right] at (4,5.25) {$\Delta M$}; 
\end{tikzpicture}
}\\
\subfloat []{
\begin{tikzpicture} [scale=0.65]
\fill [lightgray] (2,0) rectangle (2.5,6);
\draw [->] (3,-0.5) -- (3,6.5); \node [above] at (3,6.5) {$s$};
\draw [->] (-0.5,3) -- (6.5,3); \node [right] at (6.5,3) {$s_0$};
\draw [gray] (0,0) -- (0,6) -- (6,6) -- (6,0) -- (0,0);
\draw [gray] (0,0) -- (6,6);
\draw (0,0) -- (2,0) -- (2,6) -- (6,6);
\draw (2,0) -- (2.5,0) -- (2.5,6);
\node [below, fill=white] at (6,0) {$1$};
\node [below, fill=white] at (3,0) {$0$};
\node [below, fill=white] at (0,0) {$-1$};
\node [below left, fill=white] at (2,0) {$-M$}; 
\node [below, fill=white] at (2.5,0) {$-M'$}; 
\node [left, fill=white] at (0,6) {$1$};
\node [left, fill=white] at (0,3) {$0$};
\node [left, fill=white] at (0,0) {$-1$};
\draw [blue] (0,1) -- (5,6);
\draw [red] (0,0.5) -- (5.5,6);
\draw [<->] (4,5) -- (4,4.5); 
\node [right] at (4,4.25) {$\Delta M$}; 
\end{tikzpicture}
}
\subfloat []{
\begin{tikzpicture} [scale=0.65]
\draw [->] (3,-0.5) -- (3,6.5); \node [above] at (3,6.5) {$s$};
\draw [->] (-0.5,3) -- (6.5,3); \node [right] at (6.5,3) {$s_0$};
\draw [gray] (0,0) -- (0,6) -- (6,6) -- (6,0) -- (0,0);
\draw [gray] (0,1.5) -- (6,4.5);
\draw (0,1.5) -- (2.5,1.5) -- (2.5,4.5) -- (6,4.5);
\node [below, fill=white] at (6,0) {$1$};
\node [below, fill=white] at (3,0) {$0$};
\node [below, fill=white] at (0,0) {$-1$};
\node [left, fill=white] at (0,6) {$2$};
\node [left, fill=white] at (0,4.5) {$1$};
\node [left, fill=white] at (0,3) {$0$};
\node [left, fill=white] at (0,1.5) {$-1$};
\node [left, fill=white] at (0,0) {$-2$};
\draw [blue] (0,1.75) -- (6,4.75);
\draw [red] (2.5,4.5) -- (5.5,6);
\draw [red] (0,0.25) -- (2.5,1.5);
\end{tikzpicture}
}
\caption{Illustration of the quantity $s_0+M$ (blue) whose sign is $s$ (black) as a function of $s_0$, for $M>0$ (a). Area of spin flip (gray) due to a change of $M$, $\Delta M>0$ (b), $\Delta M<0$ (c), external reset (d, different $y$-axis scale).} \label{fig:spin}
\end{figure}
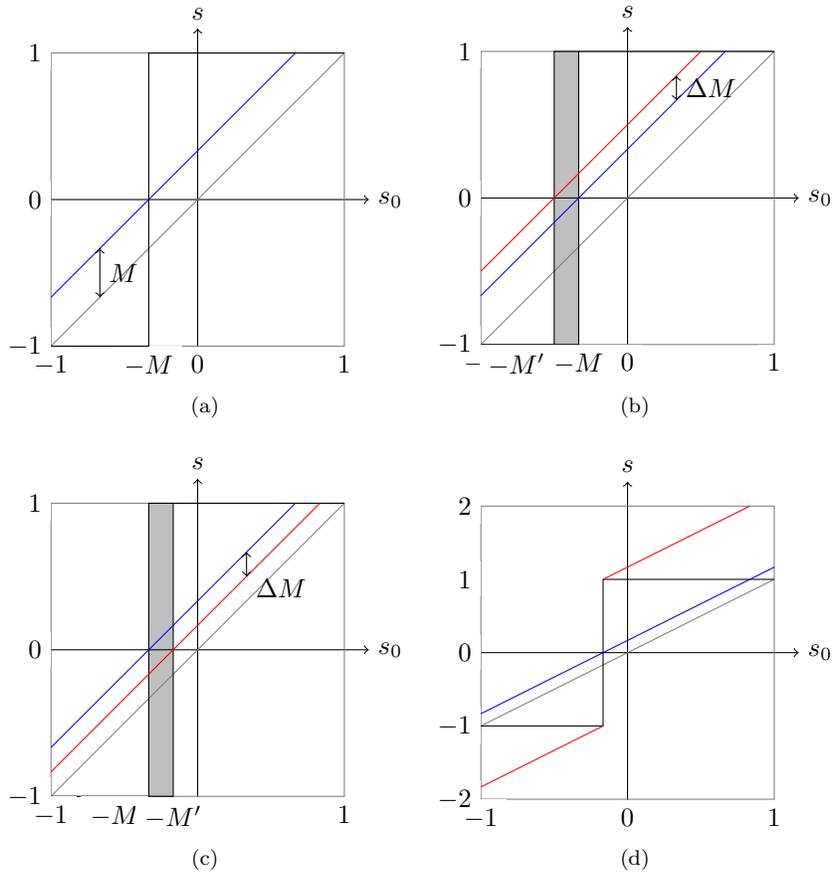

We define here the ``spin" as a binary quantity $s\in\{-1,1\}$ that varies during the particle's evolution according to the rule
\begin{equation}
s[n] = \mathrm{sign}(s_0+M[n]) \;,
\label{eqn:s_def} 
\end{equation}
see Fig.~\ref{fig:spin}a.
Clearly, the expected value of spin $\mathbb{E}[s[n]]=M[n]$. For such reason, we shall denote the variable $M$ as ``spin propensity" in the following.

Assuming that $\beta$ is constant, the magnetic force due to spin is described in analogy to the classical expression,
\begin{equation}
f_d [n]=-\mu_M \sum_{d'=1}^3 \mu_{d'} [n] \frac{\partial B_M}{\partial x_d}[n] \beta_{d'}[n]\;.
\label{eqn:magnforce}
\end{equation}
We shall assume for later use that $B_M$ is parameterizable as $\partial B_M/\partial x_d=B_F \nu_d$, with $\sum_{d=1}^3 \nu_d^2 =1$. The force is thus directed along the $\nu=\{\nu_d\}$ direction and we can define its magnitude as $f_\nu [n]=-\mu_M B_F M[n]$. 

We additionally introduce a ``magnetic energy" %
\begin{equation}
\mathcal{E}:=-\mu_MB_M \tred{\lambda} s
\;, \label{eqn:magnet}    
\end{equation} 
\tred{where $\lambda$ is a binary quantity ($\pm 1$) associated to the direction pointed by the polarization.}
By virtue of this definition, when spin changes (reverses), magnetic energy reverses as well.
Since $s_0$ is a constant for each particle, (\ref{eqn:s_def}) shows that spin flips may occur only when the spin propensity changes, that is, when $\Delta M[n]=M[n+1]-M[n]\neq 0$.
For positive (negative) $\Delta M$, a spin flip from -1 (+1) to +1 (-1) occurs with probability $\Delta M/2$ ($-\Delta M/2$).
In both cases the expected value of spin variation is equal to $\Delta M$, provided that $-1-M\leq \Delta M \leq 1-M$. 

We therefore define a ``spin-flip energy" that is attributed to particles in correspondence to any variation $\Delta M$, as the expected value of magnetic energy variation,
\begin{equation}
\delta\mathcal{E}[n]:= -\mu_MB_M \tred{\lambda}\Delta M[n]\;.
\label{eqn:denma}
\end{equation}
%

Magnetic force (\ref{eqn:magnforce}) and spin-flip energy (\ref{eqn:denma}) affect momentum and thus position of particles. In this section we shall consider a stochastic model for particle's \textit{expected} motion, that is
\begin{equation}
    x_d[n+1]=x_d[n]+v_d[n]+
    \left(\frac{1-v_d^2[n]}{2}\right)^{1/2}
    \xi[n]\;, \quad x_d[n_0]=x_{0d}\;, \label{eqn:x}
\end{equation}
\begin{equation}
    v_d[n+1]=v_d[n]\left(1+\frac{2\delta \mathcal{E}[n]}{v^2[n]}\right)^{1/2}
    +f_d[n]
    \;, \quad v_d[n_0]=v_{0d}\;, \label{eqn:v}
\end{equation}
where $x$ denotes now the expected position (a real-valued quantity rather than an integer), 
$v=\{v_d\}\in[-1,1]^3$ is denoted as ``momentum propensity", $v_0$ is its initial value that is randomly attributed at the source,
and $\xi$ represents a sequence of standard normal random variables. 
In Appendix~\ref{sec:summ}, these rules are derived from the underlying mechanism where particles move on the lattice and position takes only discrete values.
More general rules that enable the emergence of quantum behavior in position and momentum space are also summarized in that appendix.

In addition to motion rules, each time an external force is experienced, the particle undergoes an External Reset (ER) of its polarization,
\begin{equation}
\mu_d \xLeftarrow [\mathrm{ER}]{} s \beta_d \;.
\label{eqn:88}
\end{equation}
According to its definition (\ref{eqn:momprop}), the spin propensity consequently jumps to the current value of spin 
\begin{equation}
M \xLeftarrow [\mathrm{ER}]{} s\sum_{d=1}^3\beta_d^2=s  \;, \label{eqn:er_M}
\end{equation}
Note that this jump of spin propensity does not induce spin flips, see Fig.~\ref{fig:spin}d, thus there is no spin-flip energy associated to an ER.

\subsection{Probability densities}

We aim now at evaluating the pmf $\rho(s)$, which results from the particular preparation at the source and the nature of the magnetic field experienced by the ensemble of particles. We shall consider first a preparation (``pure state") for which the source polarization has a definite value $\mu_{0}$ for all the particles of the ensemble.

\subsubsection{Homogeneous field}
If the magnetic field $B_d=B_M \beta_d$ is homogeneous in space (though possibly variable with time), no magnetic force is experienced, thus no external reset occurs.
If the field is also constant, $M$ does not change with the iterations and thus is always equal to its initial value,  $M[n]\equiv M_0=\sum_{d} \mu_{0d} \beta_d$. 

From (\ref{eqn:s_def}), we have that $s$ is also constant as
\begin{equation}
s[n] = \left\{ \begin{array}{ll} 
1\:, \quad & s_0 > -M_0 \\
-1\:, \quad & s_0 < -M_0 \\
\end{array} \right. \;.
\label{eqn:ss0}
\end{equation}
Since $s_0=\mathcal{U}[-1,1]$, the probabilities of spins up and down are evaluated as 
\begin{equation}
\rho(\pm 1) = \mathbb{P}(s=\pm 1)=\frac{1\pm M_0}{2}=
\frac{1\pm\cos⁡(\mu_0,\beta)}{2}\;,
\label{eqn:rho_up}
\end{equation}
and is easily generalized to the case of a variable field, in perfect agreement with QM prediction. 

The meaning of the polarizations in the model can be now clarified. If the field is along one particular direction $d$, then $\beta_d=1$, and consequently $\rho(1)=(1+\mu_d)/2$, $\rho(-1)=(1-\mu_d)/2$. The expectation of the spin is therefore evaluated as $(1)(1+\mu_d )/2+(-1)(1+\mu_d)/2=\mu_d$. Thus the $d$-polarization represents the standard QM quantity $\langle S_d \rangle$, that is, the expected value of the spin measured along the $d$ direction.

It should be also apparent that the standard QM spinor formulation of a spin state can be retrieved by defining the complex vector quantity
\begin{equation}
    \chi=\left(\sqrt{\frac{1+\mu_{3}}{2}},\frac{\mu_{1}-\iota\mu_{2}}{\sqrt{2} \sqrt{(1+\mu_{3})}}\right)^T\;.\label{eqn:spinor}
\end{equation}
from which all standard results can be obtained.

\subsubsection{1D-inhomogeneous field (Stern-Gerlach apparatus)}
We shall consider now the case where the prepared particles pass through a Stern-Gerlach (SG) apparatus. Inside this apparatus, the field has a prevalent magnitude $B_M$ along a constant direction $\beta$ and some small inhomogeneity inducing a magnetic force of magnitude $\mu_MB_F$ along the constant direction $\nu$.

Outside the SG, $\beta=0$, $\mu\equiv\mu_0$, and so is $M=0$. Thus, spins up and down are equally distributed. 
Let us denote $n_i$ as the iteration at which particles enter the SG. The spin propensity becomes $M[n_i]=M_0$. The probability distribution of $s[n_i]$ is still given by (\ref{eqn:rho_up}). 
Due to (\ref{eqn:denma}) and (\ref{eqn:v}), the momentum is modified by the factor  $\left(1-2\mu_MB_M\tred{\lambda}M_0/v^2[n_i-1]\right)^{1/2}$.

Moreover, the presence of a magnetic force in the SG activates the External Reset condition. We shall assume, for the sake of discussion only, that the first ER occurs right after the SG entry at $n_i$. Then, $M[n_i]\Leftarrow s[n_i]$ by virtue of the ER condition (\ref{eqn:er_M}).

At the immediately next iteration, the application of (\ref{eqn:s_def}) states that $\mathbb{P}(s[n_i+1]=\sigma)=(1+\sigma s[n_i])/2$. In other words, if $s[n_i]=1$, then $s[n_i+1]$ will be 1 with probability one. Inversely, if if $s[n_i]=-1$, then $s[n_i+1]$ will surely remain -1.
Thus $\delta\mathcal{E}=0$ as discussed above after (\ref{eqn:er_M}).

At successive ER's, the situation does not change and the spin remains constant throughout the whole SG apparatus. The probability of having spins up or down, respectively, at the SG exit is thus still given by (\ref{eqn:rho_up}),
\begin{equation}
    \rho(s) = \frac{1+s\cos(\mu_0,\beta)}{2}\;.
\end{equation}
%

\subsubsection{Cascade of SG's} \label{sec:cascade}
In textbook descriptions of spin, two or more SG apparatuses in series are often employed to illustrate its non-classical properties.

In the proposed model, particles having spin $s^{(1)}=\pm 1$ at the output of the first SG have also polarization $\mu=\pm \beta^{(1)}$. At the entry of the second SG, the spin propensity is thus $M=s^{(1)}\sum_{d} \beta_{d}^{(1)}\beta_{d}^{(2)}$. Using the result of the previous section, the probability of spins up or down at the exit of the second SG is evaluated as
\begin{equation}
    \rho(s^{(2)}|s^{(1)}) = \frac{1+s^{(2)}s^{(1)}\cos(\beta^{(1)},\beta^{(2)})}{2} \;,
\end{equation}
again in perfect agreement with standard QM calculation.

\subsection{SG simulation and numerical results} \label{sec:numer1}
We shall consider a magnetic field concentrated in a certain region of space along the beam direction $x_2$ and oriented along the $x_3$ axis, with a one-dimensional inhomogeneity along the same direction, $B=(0,0,B_1 x_3)$. Even if this field does not satisfy Maxwell equation $\nabla \cdot B=0$, we choose it to simplify the notation. In fact, the literature has shown its equivalence to any ``physical" field where the inhomogeneity is along one constant direction, provided that the two directions are exchanged \cite{hsu}.

Particles are emitted one by one with $v_{01}=v_{03}=0$, while $v_{02}$ determines the average particle speed along the propagation direction, according to rules (\ref{eqn:x})--(\ref{eqn:v}).
By virtue of the equivalence (\ref{eqn:spinor}), the initial polarizations are chosen as to represent an initial spin state $\chi_0=(\chi_1,\chi_2)$, 
\begin{equation}
\mu_{03}=\chi_1\chi^*_1-\chi_2\chi^*_2, \quad \mu_{01}=\chi_1\chi_2^*+\chi_1^*\chi_2,\quad \mu_{02}=-\iota\left(\chi_1\chi_2^*-\chi_1^*\chi_2\right)
\;,
\end{equation}
where the asterisk denotes here complex conjugation.

Ensemble results are compared with those of quantum mechanics (theoretical values) obtained by using the two-component propagator \cite{hsu,reddy}
\begin{equation}
K^{(SG)} (x,t|x_0) = \frac{1}{(2\iota t)^{3/2}} \cdot \exp⁡\left(\frac{\iota\pi(x-x_0 )^2}{2t}+\frac{\iota\pi(x_1+x_{01}) \sigma_3 \phi t}{2}-\frac{\iota\pi\phi^2 t^3}{24} \right)
\label{eqn:168}
\end{equation}
in lattice units, where $\sigma_3$ denotes here the third Pauli matrix.

The theoretically expected pdf is obtained numerically from the propagated spinor $\chi(x,t)$ as $\rho(x;t)=\chi^\dagger 
\bf{1}\chi$. This pdf is to be compared with the frequency of particle 
arrivals at ``nodes" $x$ 
after $t$ iterations of model (\ref{eqn:x})--(\ref{eqn:v}).
The theoretically expected spin density 
is obtained as $\langle S_3 \rangle(x;t)=\chi^\dagger \sigma_3 \chi$. This quantity is to be compared with its counterpart in the proposed model, obtained as the difference between the frequency of arrivals of particles with $s=1$ ($\mu_3=\beta_3=1$) and of those with $s=-1$ ($\mu_3=-\beta_3=-1$).

\begin{figure}[h]
\input{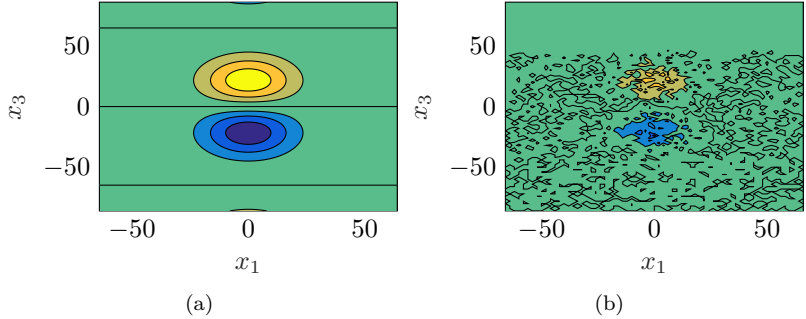}
\caption{Differential frequency of arrival (b) and theoretical spin density $\langle S_3 \rangle$ (a) for $N_p=10000$, $t=64$ as a function of position (Stern-Gerlach, Gaussian wave, $\mu_M B_F=0.1/\pi^2$,  
$\chi_1=\chi_2=1/\sqrt{2}$).}
\label{fig:sg1}
\end{figure}

\begin{figure}[h]
\centering
\input{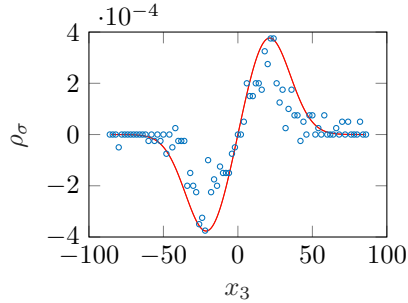}
\caption{Differential frequency of arrival (blue) and theoretical spin density (red) for $N_p=10000$, $t=64$ as a function of the inhomogeneity direction (Stern-Gerlach, Gaussian wave, $\mu_M B_F=0.1/\pi^2$, 
$\chi_1=\chi_2=1/\sqrt{2}$).}
\label{fig:sg2}
\end{figure}

Figures (\ref{fig:sg1})--(\ref{fig:sg2}) show the calculated spin density after $t=64$ iterations for a source scenario with $\mu_MB_F=0.1/\pi^2$, $\chi_1=\chi_2=1/\sqrt{2}$ (that is, $\mu_{01}=1$ in the proposed model). Globally, these result match the theoretical values, which clearly show the ``textbook" spin separation occurring along the inhomogeneity direction.


\section{Spin Entanglement} \label{sec:entang}

In this section we extend the proposed model to an ensemble of emissions of entangled particles.

\subsection{Particle emission}
Entangled particles are emitted at sources as pairs ($n_R=2$) and denoted with a superscript $R\in\{I,II\}$. 
In addition to assigning entangled momenta ($v_0^\I+v_0^\II=0$), the source preparation attributes anti-correlated entangled polarizations and spins, according to the rules
\begin{equation} \label{eqn:entanrule}
    \mu_0^{(I)}+\mu_0^{(II)}=0, \quad s_0^{(I)}+s_0^{(II)}=0\;.
\end{equation}
We shall denote, without loss of generality, $\mu_0^\I:=\mu_0$ and $s_0^\I:=s_0$, with $\sum_d\mu_{0d}^2=1$ and $s_0=\mathcal{U}[-1,1]$ as for $n_R=1$.

\subsection{Microscopic motion} 
All rules described above remain the same in the case of entangled particles, except for the fact that the quantity
\begin{equation}
  \tilde{M}^\RR:=\mathrm{sign}(M^\RR)|M^\RR|^{n_R}    \label{eqn:mtilde}
\end{equation}
replaces now $M^\RR=\sum_d \mu_d^\RR \beta_d^\RR$ in the spin dynamics (\ref{eqn:s_def}), which is rewritten as
\begin{equation}
    s^\RR[n] = \mathrm{sign}\left(s_0^\RR+\tilde{M}^\RR[n]\right)\;, \label{eqn:sR} 
\end{equation}
as well as in the magnetic force expression (\ref{eqn:magnforce}) and in the spin-flip energy expression (\ref{eqn:denma}), which are generalized accordingly.
By virtue of definition (\ref{eqn:mtilde}), when across an external reset $M^\RR$ jumps to $s^\RR$, also $\tilde{M}^\RR$ jumps to $s^\RR$.

We finally note that for $n_R=1$, equations of Sect.~\ref{sec:spin} are retrieved.

\subsection{Probability densities} \label{sec:ent_prob}

In this section we shall \tred{consider a fully random preparation (``mixed state") of the source polarization $\mu_0$. We shall} evaluate the joint pmf $\rho(s^\I,s^\II)$, representing the probability that two entangled particles arrive at either of two ``detectors" opportunely placed downstream of their respective SG apparatuses in order to intercept the beams with $s^\RR=\pm 1$. 
We start with noting that, at least in experiments with entangled photons, Bell correlations and all related statistics are obtained by counting the coincidences in arrivals at detectors. For this purpose, a data analysis procedure is required to group particles in pairs according to their arrival times, often using a time-coincidence window. A very thorough and enlightening discussion on this point can be found in \cite{larsson,michielsen,deraedt}. 
Also so-called time-tagged or coincidence loophole-free experiments eventually need some procedure to correctly identify arrivals and compute correlations, which are based on some signal generated at arrivals \cite{aguero,giustina,shalm,deraedt2017}.

Here we shall define coincidences operationally based on arrival time but more fundamentally based on arrival energy (propensity).
If we assume, without loss of generality, that the settings of the two stations are identical, these two conditions are in fact equivalent. 
Indeed, the arrival times are solely determined by the momentum acquired by the two ensembles of particles in their respective directions of propagation. By virtue of (\ref{eqn:v}), the latter are evaluated as
\begin{equation}
    v_\nu^\RR=v_{0\nu}\left(1-\frac{2\mu_MB_M\tred{\lambda}\tilde{M}_0^\RR}{v_0^2}\right)^{1/2}\;. \label{eqn:alpharr}
\end{equation}
While the initial momentum and the other parameters of influence are common to both ensembles, it is therefore the cosine term
$M_0^\RR=\cos⁡(\mu_0^\RR,\beta^\RR)$ that controls the momentum through (\ref{eqn:alpharr}) and ultimately 
defines energy propensity and time of arrival at the respective detectors.

For this reason, a coincidence in arrivals at the detectors is \textit{expectedly} recorded when the two particles have the same value of $M_0$, that is, when
\begin{equation}
M_0^\I=\sum_{d=1}^3 \beta_d^\I \mu_{0d}^\RR = \sum_{d} \beta_d^\II \mu_{0d}^\RR=M_0^\II
\label{eqn:smu}
\end{equation}
It is now easy to show that, for each selection of $\beta^\I$, $\beta^\II$, two values of $M_0$ fulfill (\ref{eqn:smu}), namely,
\begin{equation}
\hat{M}_0
=\pm \sqrt{\frac{1-\sum_{d=1}^3 \beta^\I_d \beta^\II_d}{2}} \;, 
\label{eqn:smu1}
\end{equation}
having opposite signs and equal probability. Geometrically, these values correspond to the two unit $\mu_0$ vectors bisecting the angle between the two directions $\beta^\I$, $\beta^\II$ and co-planar to the same, that is 
$$\angle\hat{\mu}_0=(\angle\beta^\I+\angle\beta^\II)/2\;.$$
All other values of $\mu_0$ give rise to coincidences only with a very small probability and thus do not contribute to the joint pmf.

Similarly to the non-entangled case, inside the SG $s^\RR$ remains constant at the entry value $\mathrm{sign}(s_0^\RR+\mathrm{sign}(M_0)M_0^2)$.
The distributions of $s^\RR$ are therefore found as
\begin{equation}
s^\I = \left\{ \begin{array}{ll} 
1\;, \quad & s_0 > -\mathrm{sign}(M_0)M_0^2 \\
-1\;, \quad & s_0 < -\mathrm{sign}(M_0)M_0^2 
\end{array} \right. \label{eqn:sI}
\end{equation}
and
\begin{equation}
s^\II = \left\{ \begin{array}{ll} 
1\;, \quad & s_0 < \mathrm{sign}(M_0)M_0^2 \\
-1\;, \quad & s_0 > \mathrm{sign}(M_0)M_0^2 
\end{array} \right. \label{eqn:sII}
\end{equation}

\begin{figure}
\centering
\subfloat []{
\begin{tikzpicture} [scale=0.65]

\draw [->] (3,-0.5) -- (3,6.5); 
\node [above] at (3,6.5) {$s$};
\draw [->] (-0.5,3) -- (6.5,3); 
\node [right] at (6.5,3) {$s_0$};
\draw [gray] (0,0) -- (0,6) -- (6,6) -- (6,0) -- (0,0);
\draw [gray] (0,0) -- (6,6);
\draw [gray] (0,6) -- (6,0);

\draw [dashed] (2,0) -- (2,6);
\draw [dashed] (4,0) -- (4,6);

\node [below, fill=white] at (6,0) {$1$};
\node [below, fill=white] at (4,0) {$M_0^2$}; \node [below, fill=white] at (2,0) {$-M_0^2$}; \node [below, fill=white] at (0,0) {$-1$};

\draw [blue] (0,1) -- (5,6);
\draw [red] (1,6) -- (6,1);
\draw [<->] (1,2) -- (1,1);
\node [right] at (1,1.5) {$M_0^2$}; 
\end{tikzpicture}
}
\subfloat []{
\begin{tikzpicture} [scale=0.65]

\draw [->] (3,-0.5) -- (3,6.5); 
\node [above] at (3,6.5) {$s$};
\draw [->] (-0.5,3) -- (6.5,3); 
\node [right] at (6.5,3) {$s_0$};
\draw [gray] (0,0) -- (0,6) -- (6,6) -- (6,0) -- (0,0);
\draw [gray] (0,0) -- (6,6);
\draw [gray] (0,6) -- (6,0);

\draw [dashed] (2,0) -- (2,6);
\draw [dashed] (4,0) -- (4,6);

\node [below, fill=white] at (6,0) {$1$};
\node [below, fill=white] at (4,0) {$M_0^2$};
\node [below, fill=white] at (2,0) {$-M_0^2$}; \node [below, fill=white] at (0,0) {$-1$};

\draw [blue] (1,0) -- (6,5);
\draw [red] (0,5) -- (5,0);
\draw [<->] (5,4) -- (5,5); \node [left] at (5,4.5) {$M_0^2$};

\end{tikzpicture}
}
\caption{Illustration of the quantities $s_0^\I+\mathrm{sign}(M_0)M_0^2$ whose sign is $s^\I$ (blue) and $s_0^\II+\mathrm{sign}(M_0)M_0^2$ whose sign is $s^\II$ (red), as a function of $s_0$, for $M_0>0$ (a) and $M_0<0$ (b).} \label{fig:spin2}
\end{figure}
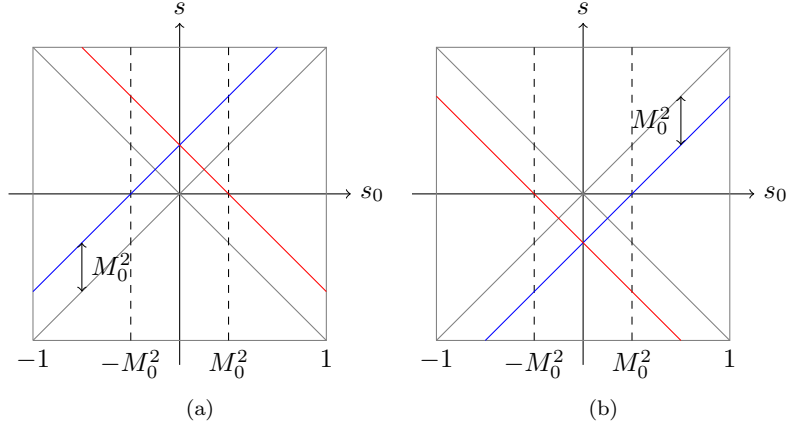

With the help of Fig. \ref{fig:spin2}, we further note that $\{s^\I,s^\II\}=\{1,1\}$ for
$s_0\in[-M_0^2,M_0^2]$ if $M_0>0$.
Likewise, $\{s^\I,s^\II\}=\{-1,-1\}$ 
for $s_0\in[-M_0^2,M_0^2]$ if $M_0<0$.
Regardless of the sign of $M_0$, $\{s^\I,s^\II\}=\{-1,1\}$ for $s_0<-M_0^2$
and $\{s^\I,s^\II\}=\{1,-1\}$ for
$s_0 > M_0^2$.

Finally, recalling that $\rho(s_0)=1/2$, the joint distribution is evaluated as
\begin{eqnarray}
&& \rho(1,1) = 
\mathbb{P}(M_0>0)\cdot\frac{2M_0^2}{2}=\frac{M_0^2}{2}\\ 
&& \rho(-1,-1) = \mathbb{P}(M_0<0)\cdot\frac{2M_0^2}{2}=\frac{M_0^2}{2}\\ 
&& \rho(1,-1) = \frac{1-M_0^2}{2} = \rho(-1,1)\;.
\end{eqnarray}
Since from (\ref{eqn:smu1}) the only meaningful value is $\hat{M}_0^2=(1-\cos(\beta^\I,\beta^\II))/2$, the joint distribution reads
\begin{equation}
\rho(s^\I,s^\II) = \frac{1-s^\I s^\II\cos(\beta^\I,\beta^\II)}{4}\;, \label{eqn:rhosigma}
\end{equation}
and the expected value of the product $s^\I s^\II$ is $-\cos(\beta^\I,\beta^\II)$,
that is, precisely the QM prediction. 

\tred{In order to obtain this result, a key role is played by the first of rules (\ref{eqn:entanrule}) that enforces full correlation of the two emissions. In the case of a totally uncorrelated preparation where $\mu_0^\I$ and $\mu_0^\II$ are independent random variables, the result (\ref{eqn:rhosigma}) becomes $\rho(s^\I,s^\II)=1/4$, independent of the SG orientations, and $\mathbb{E}(s^\I s^\II)=0$. A classically correlated result with $\mathbb{E}(s^\I s^\II)=-1/2\cos(\beta^\I,\beta^\II)$ is obtained for a statistical (50\%-50\%) mixture of the aforementioned preparations.}

\tred{Indeed, similarly to (\ref{eqn:spinor}), also for two-particle systems it is possible to establish a correspondence between the preparation rule in our model and the QM state with its density matrix.}
\tgreen{Detailed calculations of these cases are not shown here.}

\subsection{Numerical Results} \label{sec:numer}
We aim at representing here a textbook two-channel Bell test experiment. A source produces pairs of entangled particles, sent in opposite directions. Each particle beam encounters a SG. Emerging particles from each channel are detected and coincidences in arrivals counted.
Similarly to the non-entangled scenario simulated in Sect.~\ref{sec:numer1}, we shall take $\beta^\RR=\nu^\RR$, i.e., an inhomogeneity directed along the field in both SG. While the orientation $\beta^\II$ is fixed, $\beta^\I$ is varied between $-\pi$ and $\pi$ in the plane $x_1$--$x_3$. The direction of the two emitted beams is taken as $\pm x_2$. 
Arrivals are registered when particles' position $x_2=\ell$.
Coincidences are then registered when arrival times do not differ by more than a coincidence window $W$.

In the proposed model, particles are emitted at the respective sources with $v_{01}=v_{03}=0$, while $v_{02}$ would determine the average particle speed along the propagation direction, according to rules (\ref{eqn:x})--(\ref{eqn:v}).
The initial polarizations are randomly chosen. \tred{In total, $N_p$ emissions are simulated.}
%

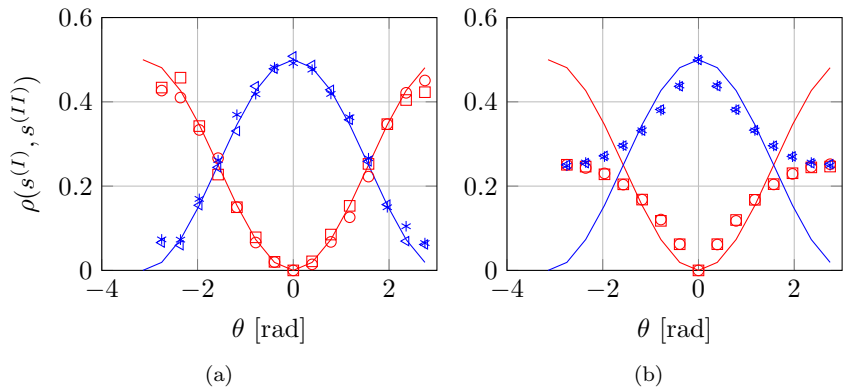
\begin{figure}[ht]
\centering
\subfloat[]{\label{fig:bell_a}
%
%
\definecolor{mycolor1}{rgb}{0.00000,0.44700,0.74100}%
\begin{tikzpicture}

\begin{axis}[%
width=4.521in,
height=3.566in,
at={(0.758in,0.481in)},
scale=0.45,
xmin=-4,
xmax=3,
xlabel={\tred{$\theta$} [rad]},
ymin=0,
ymax=0.6,
ylabel={$\rho(s^\I,s^\II)$},
axis background/.style={fill=white},
xmajorgrids,
ymajorgrids,
legend style={legend cell align=left, align=left, draw=white!15!black}
]
\addplot [color=red, draw=none, mark=o, mark options={solid, red}]
  table[row sep=crcr]{%
-2.74889357189107	0.42668863261944\\
-2.35619449019234	0.410112359550562\\
-1.96349540849362	0.333333333333333\\
-1.5707963267949	0.266355140186916\\
-1.17809724509617	0.149922720247295\\
-0.785398163397448	0.0661434977578475\\
-0.392699081698724	0.019680196801968\\
0	0\\
0.392699081698724	0.0143283582089552\\
0.785398163397448	0.0676117775354417\\
1.17809724509617	0.126126126126126\\
1.5707963267949	0.222556390977444\\
1.96349540849362	0.347003154574132\\
2.35619449019234	0.421412300683371\\
2.74889357189107	0.450681198910082\\
};

\addplot [color=red, draw=none, mark=square, mark options={solid, red}]
  table[row sep=crcr]{%
-2.74889357189107	0.43382756727073\\
-2.35619449019234	0.457303370786517\\
-1.96349540849362	0.342424242424242\\
-1.5707963267949	0.227414330218069\\
-1.17809724509617	0.149922720247295\\
-0.785398163397448	0.07847533632287\\
-0.392699081698724	0.0202952029520295\\
0	0\\
0.392699081698724	0.0214925373134328\\
0.785398163397448	0.0850599781897492\\
1.17809724509617	0.153153153153153\\
1.5707963267949	0.252631578947368\\
1.96349540849362	0.347003154574132\\
2.35619449019234	0.404328018223235\\
2.74889357189107	0.422888283378747\\
};

\addplot [color=blue, draw=none, mark=asterisk, mark options={solid, blue}]
  table[row sep=crcr]{%
-2.74889357189107	0.0735859417902252\\
-2.35619449019234	0.0730337078651685\\
-1.96349540849362	0.16969696969697\\
-1.5707963267949	0.261682242990654\\
-1.17809724509617	0.369397217928903\\
-0.785398163397448	0.418161434977578\\
-0.392699081698724	0.481549815498155\\
0	0.491822429906542\\
0.392699081698724	0.476417910447761\\
0.785398163397448	0.418756815703381\\
1.17809724509617	0.363363363363363\\
1.5707963267949	0.266165413533835\\
1.96349540849362	0.149842271293375\\
2.35619449019234	0.104783599088838\\
2.74889357189107	0.064850136239782\\
};

\addplot [color=blue, draw=none, mark=triangle, mark options={solid, rotate=90, blue}]
  table[row sep=crcr]{%
-2.74889357189107	0.0658978583196046\\
-2.35619449019234	0.0595505617977528\\
-1.96349540849362	0.154545454545455\\
-1.5707963267949	0.244548286604361\\
-1.17809724509617	0.330757341576507\\
-0.785398163397448	0.437219730941704\\
-0.392699081698724	0.478474784747847\\
0	0.508177570093458\\
0.392699081698724	0.487761194029851\\
0.785398163397448	0.428571428571429\\
1.17809724509617	0.357357357357357\\
1.5707963267949	0.258646616541353\\
1.96349540849362	0.15615141955836\\
2.35619449019234	0.0694760820045558\\
2.74889357189107	0.0615803814713896\\
};

\addplot [color=red]
  table[row sep=crcr]{%
-3.14159265358979	0.5\\
-2.74889357189107	0.480969883127822\\
-2.35619449019234	0.426776695296637\\
-1.96349540849362	0.345670858091272\\
-1.5707963267949	0.25\\
-1.17809724509617	0.154329141908728\\
-0.785398163397448	0.0732233047033631\\
-0.392699081698724	0.0190301168721783\\
0	0\\
0.392699081698724	0.0190301168721783\\
0.785398163397448	0.0732233047033631\\
1.17809724509617	0.154329141908727\\
1.5707963267949	0.25\\
1.96349540849362	0.345670858091273\\
2.35619449019234	0.426776695296637\\
2.74889357189107	0.480969883127822\\
};

\addplot [color=blue]
  table[row sep=crcr]{%
-3.14159265358979	1.87469972832732e-33\\
-2.74889357189107	0.0190301168721783\\
-2.35619449019234	0.0732233047033631\\
-1.96349540849362	0.154329141908728\\
-1.5707963267949	0.25\\
-1.17809724509617	0.345670858091272\\
-0.785398163397448	0.426776695296637\\
-0.392699081698724	0.480969883127822\\
0	0.5\\
0.392699081698724	0.480969883127822\\
0.785398163397448	0.426776695296637\\
1.17809724509617	0.345670858091273\\
1.5707963267949	0.25\\
1.96349540849362	0.154329141908727\\
2.35619449019234	0.0732233047033631\\
2.74889357189107	0.0190301168721784\\
};


\end{axis}
\end{tikzpicture}%
}
\subfloat[]{\label{fig:bell_b}
%
%
\begin{tikzpicture}

\begin{axis}[%
width=4.521in,
height=3.566in,
at={(0.758in,0.481in)},
scale=0.45,
xmin=-4,
xmax=3,
xlabel={\tred{$\theta$} [rad]},
ymin=0,
ymax=0.6,
axis background/.style={fill=white},
xmajorgrids,
ymajorgrids,
legend style={legend cell align=left, align=left, draw=white!15!black}
]
\addplot [color=red, draw=none, mark=o, mark options={solid, red}]
  table[row sep=crcr]{%
-2.74889357189107	0.25075\\
-2.35619449019234	0.24307\\
-1.96349540849362	0.23017\\
-1.5707963267949	0.20349\\
-1.17809724509617	0.1679\\
-0.785398163397448	0.12067\\
-0.392699081698724	0.0623974712919617\\
0	0\\
0.392699081698724	0.0614178111652379\\
0.785398163397448	0.11752\\
1.17809724509617	0.16757\\
1.5707963267949	0.20395\\
1.96349540849362	0.22905\\
2.35619449019234	0.24534\\
2.74889357189107	0.25197\\
};

\addplot [color=red, draw=none, mark=square, mark options={solid, red}]
  table[row sep=crcr]{%
-2.74889357189107	0.25064\\
-2.35619449019234	0.24625\\
-1.96349540849362	0.22808\\
-1.5707963267949	0.20427\\
-1.17809724509617	0.16804\\
-0.785398163397448	0.11762\\
-0.392699081698724	0.0620773816668667\\
0	0\\
0.392699081698724	0.0622180432325374\\
0.785398163397448	0.11961\\
1.17809724509617	0.16751\\
1.5707963267949	0.20477\\
1.96349540849362	0.23059\\
2.35619449019234	0.24448\\
2.74889357189107	0.24643\\
};

\addplot [color=blue, draw=none, mark=asterisk, mark options={solid, blue}]
  table[row sep=crcr]{%
-2.74889357189107	0.24835\\
-2.35619449019234	0.25605\\
-1.96349540849362	0.27007\\
-1.5707963267949	0.29601\\
-1.17809724509617	0.33247\\
-0.785398163397448	0.37985\\
-0.392699081698724	0.438662825591166\\
0	0.500707929847769\\
0.392699081698724	0.439177361434816\\
0.785398163397448	0.38146\\
1.17809724509617	0.33212\\
1.5707963267949	0.29541\\
1.96349540849362	0.26977\\
2.35619449019234	0.25597\\
2.74889357189107	0.24961\\
};

\addplot [color=blue, draw=none, mark=triangle, mark options={solid, rotate=90, blue}]
  table[row sep=crcr]{%
-2.74889357189107	0.25026\\
-2.35619449019234	0.25463\\
-1.96349540849362	0.27168\\
-1.5707963267949	0.29623\\
-1.17809724509617	0.33159\\
-0.785398163397448	0.38186\\
-0.392699081698724	0.436862321450006\\
0	0.499292070152231\\
0.392699081698724	0.437186784167409\\
0.785398163397448	0.38141\\
1.17809724509617	0.3328\\
1.5707963267949	0.29587\\
1.96349540849362	0.27059\\
2.35619449019234	0.25421\\
2.74889357189107	0.25199\\
};

\addplot [color=red]
  table[row sep=crcr]{%
-3.14159265358979	0.5\\
-2.74889357189107	0.480969883127822\\
-2.35619449019234	0.426776695296637\\
-1.96349540849362	0.345670858091272\\
-1.5707963267949	0.25\\
-1.17809724509617	0.154329141908728\\
-0.785398163397448	0.0732233047033631\\
-0.392699081698724	0.0190301168721783\\
0	0\\
0.392699081698724	0.0190301168721783\\
0.785398163397448	0.0732233047033631\\
1.17809724509617	0.154329141908727\\
1.5707963267949	0.25\\
1.96349540849362	0.345670858091273\\
2.35619449019234	0.426776695296637\\
2.74889357189107	0.480969883127822\\
};

\addplot [color=blue]
  table[row sep=crcr]{%
-3.14159265358979	1.87469972832732e-33\\
-2.74889357189107	0.0190301168721783\\
-2.35619449019234	0.0732233047033631\\
-1.96349540849362	0.154329141908728\\
-1.5707963267949	0.25\\
-1.17809724509617	0.345670858091272\\
-0.785398163397448	0.426776695296637\\
-0.392699081698724	0.480969883127822\\
0	0.5\\
0.392699081698724	0.480969883127822\\
0.785398163397448	0.426776695296637\\
1.17809724509617	0.345670858091273\\
1.5707963267949	0.25\\
1.96349540849362	0.154329141908727\\
2.35619449019234	0.0732233047033631\\
2.74889357189107	0.0190301168721784\\
};

\end{axis}
\end{tikzpicture}%
}
\caption{Frequency of coincidences of arrivals $(1,1)$ (circles), $(-1,-1)$ (squares), $(1,-1)$ (stars), and $(-1,1)$ (triangles), with their theoretical values $\rho(1,1)=\rho(-1,-1)$ (red solid) and $\rho(1,-1)=\rho(-1,1)$ (blue solid) as a function of the angle \tred{$\theta$} (rad) between $\beta^\I$ and $\beta^\II$. Bell experiment, Gaussian-wave beams, $\mu_MB_M=10^{-5}$, \tred{$N_p=1\cdot 10^5$}, 
$v_{02}=0.2$, 
\tred{$\ell=0.2\cdot 10^{11}$, $W=10^6$ (a) and $W=10^8 (b)$}.} 
\end{figure}

Ensemble results are compared with QM prediction (\ref{eqn:rhosigma}).
Figure~\ref{fig:bell_a} shows the frequency of the four types of coincidences as a function of the angular difference \tred{$\theta$} between the two fields, with $\mu_MB_M=10^{-5}$, $N_p=1\cdot 10^{\tred{5}}$, 
$v_{02}=0.2$, \tred{$\ell=0.2\cdot 10^{11}$, $W=10^6$.} When compared with the QM predictions, these results confirm the substantial equivalence of the two models as anticipated in the previous section. 
\tred{Figure~\ref{fig:bell_b} is obtained with a much larger $W=10^8$ and shows that, under these circumstances, the QM correlations are lost. However, that does not mean that the classical correlation is found, since that cannot depend on the coincidence window chosen, while particle emissions are still prepared according to the ``entangled" rule (\ref{eqn:entanrule}).}

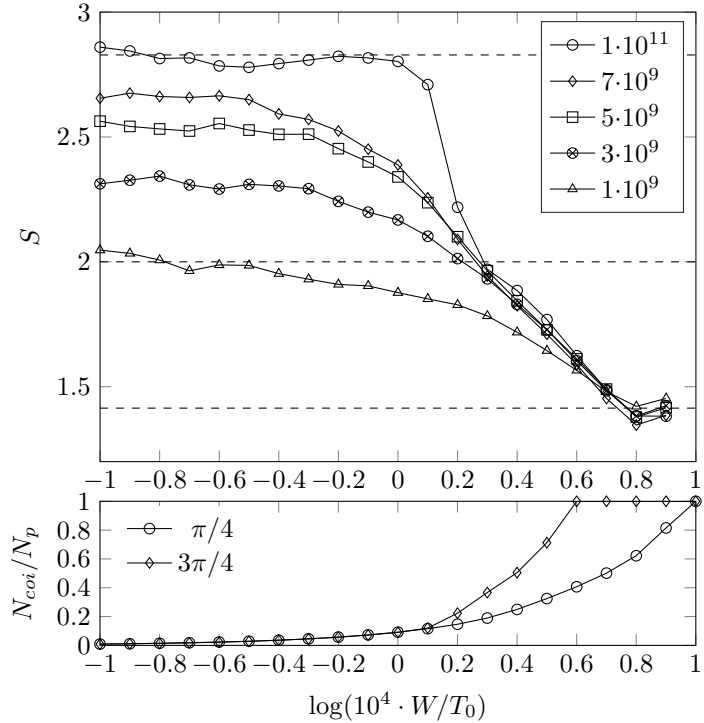
\begin{figure}[ht]
    \centering
%
%
\definecolor{mycolor1}{rgb}{0.46600,0.67400,0.18800}%
\definecolor{mycolor2}{rgb}{0.30100,0.74500,0.93300}%
\definecolor{mycolor3}{rgb}{0.00000,0.44700,0.74100}%
\definecolor{mycolor4}{rgb}{0.85000,0.32500,0.09800}%
\definecolor{mycolor5}{rgb}{0.49400,0.18400,0.55600}%
\definecolor{mycolor6}{rgb}{0.63500,0.07800,0.18400}%
\definecolor{mycolor7}{rgb}{0.92900,0.69400,0.12500}%
\begin{tikzpicture}

\begin{axis}[%
width=4.521in,
height=3.566in,
at={(0.758in,0.481in)},
scale=0.8,
xmin=-1,
xmax=1,
ymin=1.2,
ymax=3,
ylabel={$S$},
axis background/.style={fill=white},
legend style={legend cell align=left, align=left, draw=white!15!black}
]

\addplot [color=black, mark=o, mark options={solid, black}]
  table[row sep=crcr]{%
-1	2.85953817349698\\
-0.9	2.8443598021597\\
-0.8	2.81400427692206\\
-0.7	2.8166236760391\\
-0.6	2.78451751606408\\
-0.5	2.77892555205988\\
-0.4	2.7938055036795\\
-0.3	2.80771132822142\\
-0.2	2.82303226069795\\
-0.0999999999999996	2.81631184713447\\
0	2.80245813205052\\
0.0999999999999996	2.70953599779809\\
0.2	2.21860830923504\\
0.3	1.96990856560408\\
0.4	1.88491203556779\\
0.5	1.76880346873951\\
0.6	1.6241876870174\\
0.7	1.48804445284821\\
0.8	1.38277664898505\\
0.9	1.38185900747707\\
};
\addlegendentry{1$\cdot 10^{11}$}

\addplot [color=black, mark=diamond, mark options={solid, black}]
  table[row sep=crcr]{%
-1	2.65507097377183\\
-0.9	2.67537406409815\\
-0.8	2.66202455825097\\
-0.7	2.65819738614893\\
-0.6	2.66474662803373\\
-0.5	2.64953901704104\\
-0.4	2.59224449991465\\
-0.3	2.57019153322988\\
-0.2	2.52395802419885\\
-0.0999999999999996	2.45107471763851\\
0	2.38814472586968\\
0.0999999999999996	2.25628262521959\\
0.2	2.08962781358373\\
0.3	1.94318688348344\\
0.4	1.82434821282749\\
0.5	1.71001178447959\\
0.6	1.58843857508624\\
0.7	1.45284377363595\\
0.8	1.3463291247702\\
0.9	1.38555450234227\\
};
\addlegendentry{7$\cdot$10$^9$}

\addplot [color=black, mark=square, mark options={solid, black}]
  table[row sep=crcr]{%
-1	2.56323684835773\\
-0.9	2.54221650582792\\
-0.8	2.53213883965651\\
-0.7	2.52388840754797\\
-0.6	2.55386864597391\\
-0.5	2.52766900633898\\
-0.4	2.51036938965126\\
-0.3	2.51118395445173\\
-0.2	2.45316413688404\\
-0.0999999999999996	2.3998841404403\\
0	2.34001613931344\\
0.0999999999999996	2.23693910028489\\
0.2	2.10036092489136\\
0.3	1.96446766002537\\
0.4	1.84420873005844\\
0.5	1.72790584766471\\
0.6	1.61165275991914\\
0.7	1.49053315250963\\
0.8	1.37852836093752\\
0.9	1.41564148442581\\
};
\addlegendentry{5$\cdot$10$^9$}

\addplot [color=black, mark=otimes, mark options={solid, black}]
  table[row sep=crcr]{%
-1	2.31239154013015\\
-0.9	2.32672473143421\\
-0.8	2.34315373208107\\
-0.7	2.30827845870384\\
-0.6	2.29145250458022\\
-0.5	2.31024780142398\\
-0.4	2.30387562781546\\
-0.3	2.29278066901016\\
-0.2	2.24209492765342\\
-0.0999999999999996	2.19894384260395\\
0	2.1677010874232\\
0.0999999999999996	2.10227482879082\\
0.2	2.01296425166459\\
0.3	1.93196819037062\\
0.4	1.82947610997648\\
0.5	1.72802659798272\\
0.6	1.60420923064658\\
0.7	1.48358733500008\\
0.8	1.38299351321464\\
0.9	1.42389914050858\\
};
\addlegendentry{3$\cdot$10$^9$}

\addplot [color=black, mark=triangle, mark options={solid, black}]
  table[row sep=crcr]{%
-1	2.04696660299058\\
-0.9	2.03381928400162\\
-0.8	2.00710292329132\\
-0.7	1.96396483159237\\
-0.6	1.9876080031131\\
-0.5	1.98566359855572\\
-0.4	1.95271651637101\\
-0.3	1.93040435299132\\
-0.2	1.90959485009454\\
-0.0999999999999996	1.90372285253853\\
0	1.87647090277291\\
0.0999999999999996	1.85168424471102\\
0.2	1.82786293590684\\
0.3	1.78343146899806\\
0.4	1.71800637327427\\
0.5	1.64438068891363\\
0.6	1.56559713873063\\
0.7	1.4803209282602\\
0.8	1.42119592406289\\
0.9	1.45297501313747\\
};
\addlegendentry{1$\cdot$10$^9$}

\addplot [color=black, dashed]
  table[row sep=crcr]{%
-1	2.82842712474619\\
1	2.82842712474619\\
};

\addplot [color=black, dashed]
  table[row sep=crcr]{%
-1	2\\
1	2\\
};

\addplot [color=black, dashed]
  table[row sep=crcr]{%
-1	1.4142\\
1	1.4142\\
};

\end{axis}

\begin{axis}[%
width=4.521in,
height=1.566in,
at={(0.758in,-0.481in)},
scale=0.8,
xmin=-1,
xmax=1,
xlabel={$\log(10^4\cdot W/T_0)$},
ymin=0,
ymax=1,
ylabel={$N_{coi}/N_p$},
axis background/.style={fill=white},
legend pos = north west,
legend style={legend cell align=right, align=right, draw=none}
]

\addplot [color=black, mark=o, mark options={solid, black}]
  table[row sep=crcr]{%
-1	0.00917\\
-0.9	0.01155\\
-0.8	0.01459\\
-0.7	0.01831\\
-0.6	0.02319\\
-0.5	0.02893\\
-0.4	0.0359\\
-0.3	0.04527\\
-0.2	0.05752\\
-0.0999999999999996	0.07295\\
0	0.09182\\
0.0999999999999996	0.1157\\
0.2	0.1466\\
0.3	0.18983\\
0.4	0.2497\\
0.5	0.32479\\
0.6	0.40677\\
0.7	0.50143\\
0.8	0.62238\\
0.9	0.81494\\
1	1\\
};
\addlegendentry{$\pi/4$}

\addplot [color=black, mark=diamond, mark options={solid, black}]
  table[row sep=crcr]{%
-1	0.00878\\
-0.9	0.01105\\
-0.8	0.01436\\
-0.7	0.01785\\
-0.6	0.02233\\
-0.5	0.02785\\
-0.4	0.03595\\
-0.3	0.04509\\
-0.2	0.05686\\
-0.0999999999999996	0.07146\\
0	0.09014\\
0.0999999999999996	0.12021\\
0.2	0.22325\\
0.3	0.36659\\
0.4	0.5048\\
0.5	0.71271\\
0.6	0.99985\\
0.7	1\\
0.8	1\\
0.9	1\\
1	1\\
};
\addlegendentry{$3\pi/4$}

\end{axis}

\end{tikzpicture}%
    \caption{\tred{Top panel:} correlation parameter as a function of the coincidence window (simulation data in Fig.~\ref{fig:bell_a}) \tred{and for various mean arrival times $T_0$}. The dashed lines are the Bell limit ($2\sqrt{2}$), the HV theory limit (2) \tred{and the classical limit ($\sqrt{2}$)}. \tred{Bottom panel: Frequency of coincidences as a function of the coincidence window for $\theta=\pi/4$ and $3\pi/4$, respectively}.}
\label{fig:Scorr}
\end{figure}

Figure~\ref{fig:Scorr} shows the Bell parameter $S$ calculated for the Bell test angles \tblue{($\beta_1^\I=0,\beta_1^\II=3\pi/4,\beta_2^\I=-\pi/2,\beta_2^\II=-3\pi/4$)} 
for increasing values of the coincidence window $W$. 
\tred{Various mean arrival times $T_0:=\ell/v_{02}$ have been tested. For larger times, the relative spreading of the particle beams is smaller (by virtue of rule (\ref{eqn:x})) and the beam is more concentrated around its axis. The correlation parameter reaches the Bell limit for small values of $W/T_0$, shows a pronounced drop to the hidden variable theory limit of 2, then further decreases. For smaller values of $T_0$, the curve of $S$ is generally smoother and reaches lower maximum values.}
\tred{Overall,} the trend is similar to that experimentally observed with time coincidences \cite{aguero}, \tred{with BI violation observed for small enough coincidence windows.}

\tred{Note that the large-$W$ limit of $S$ ($2-\sqrt{2}/\pi\approx 1.55$) is slightly higher than the classical limit $\sqrt{2}\approx 1.52$ and confirms what observed in Fig.~\ref{fig:bell_b}, that particles are still entangled and not classically correlated regardless of the coincidence window chosen.}
%

\tred{The figure also shows the relative rate of coincidences as a function of $W$ for the two Bell test angle differences and $T_0=10^{11}$. Clearly, such a rate increases monotonically and reaches unity when $W$ becomes larger than the maximum difference of arrival times. For small windows the number of coincidences is sensibly the same for both test angles considered. Having considered 20 repetitions of the simulation of Fig.~\ref{fig:Scorr} ($N_p=1\cdot 10^5$), we have obtained $N_{coi}=892\pm30$ for $\theta=\pi/4$ and $N_{coi}=900\pm26.5$ for $\theta=3\pi/4$.}

\tblue{
One could argue that the method of counting the coincidences using a floating time window $W$ is still subject to the coincidence loophole and that one should use instead a fixed-time-slot (FTS) method 
in conjunction with a different Bell-type inequality
 \cite{larsson2013,aguero2012}. 
In fact, in the proposed model, coincident arrival times are highly concentrated around an average value and there is no time-shifting of coincidences.
As a consequence, using FTS yields substantially the same coincidence pairs and eventually the same results shown above.
}

%

\subsection{Discussion}

The previous sections have shown that the proposed model exactly reproduces the joint pmf of a spin-entangled quantum system where coincidences in spin are counted only if they are accompanied by coincidence in arrival at detectors, that is, the regime that Bell test experiments aim to reach \cite{deraedt}.
Nevertheless, nowhere in the proposed model, particles, say, II ``know" about which magnetic field experience particles I, thus locality applies, certainly together with realism.
At this point, the reader may wonder why, 
despite these premises, 
Bell's inequalities are violated and the QM statistics are correctly reproduced.

Within a hidden variable model, the outcomes $\sigma^\I$, $\sigma^\II$ of a Bell test experiment are functions of the settings $\beta^\I$, $\beta^\II$ of the two apparatuses and of some ‘hidden variable’ vector (HV) $\lambda\in\Lambda$, where $\Lambda$ is some (possibly, multidimensional) probability space, \textit{i.e.}, $\sigma^\I=\sigma^\I(\beta^\I,\lambda)$, $\sigma^\II=\sigma^\II(\beta^\II,\lambda)$.
The joint probability\footnote{For deterministic models, these probabilities can be represented by Dirac-delta pdf.} of the two outputs given the apparatus settings is generally expressed as
\begin{equation}
  P(\sigma^\I,\sigma^\II|\beta^\I,\beta^\II)=\int P(\sigma^\I,\sigma^\II|\beta^\I,\beta^\II,\lambda)P(\lambda|\beta^\I,\beta^\II))d\lambda\;. \label{eqn:bellint}  
\end{equation}

In this framework, Bell's inequalities can be derived from certain assumptions. Often referred to as locality assumption is the factorability condition (FC) \cite{clauser} 
\begin{equation}
P(\sigma^\I,\sigma^\II|\beta^\I,\beta^\II,\lambda)=P(\sigma^\I|\beta^\I,\lambda)P(\sigma^\II|\beta^\II,\lambda)\;,  \label{eqn:fc}
\end{equation}
which can be regarded as the conjunction of the outcome independence (OI) assumption, 
\begin{equation}
 P(\sigma^\I|\sigma^\II,\beta^\I,\beta^\II,\lambda)=P(\sigma^\I|\beta^\I,\beta^\II,\lambda)\;,   \label{eqn:oi}
\end{equation}
and the parameter independence (PI) assumption,
\begin{equation}
P(\sigma^\I|\beta^\I,\beta^\II,\lambda)=P(\sigma^\I|\beta^\I,\lambda)    \label{eqn:pi}
\end{equation}
(and similarly for $\sigma^\II$). 
The other assumption is measurement independence (MI), which says that the distribution of the hidden variables that determine the measurement outcomes is independent of the setting parameter of the apparatus,
\begin{equation}
  P(\lambda|\beta^\I,\beta^\II)=P(\lambda)\;. \label{eqn:mi}    
\end{equation}

In our model, hidden variables are $\lambda=\{s_0,\mu_0\}$, while the binary outputs can be defined as the spin outputs submitted to coincidence of arrivals within the time window $W$,
\begin{equation}
    \sigma^\RR:=[\delta(s^\RR-1)-\delta(s^\RR+1)]\Theta(W-|\Delta T|)\;,
\end{equation}
where $\Theta$ denotes the unit step function, $\delta$ the Dirac pulse function, and $T$ the arrival time.
By letting $W\rightarrow 0$, conditional pdf's are obtained as
\begin{align}
P(\sigma^\I|\sigma^\II,\beta,\lambda)&=\delta(\sigma^\I-\mathrm{sign}(s_0+\tilde{M}_0^\I))\rho_G(\tilde{M}_0^\I-\tilde{M}_0^\II)\;,    \\ 
 P(\sigma^\II|\sigma^\I,\beta,\lambda)&=\delta(\sigma^\II-\mathrm{sign}(-s_0+\tilde{M}_0^\II))\rho_G(\tilde{M}_0^\I-\tilde{M}_0^\II) \;,   
\end{align}
where $\rho_G$ is some bell-shaped distribution with zero mean that represents the dispersion of arrival times around the expected value resulting from momentum (\ref{eqn:alpharr}) and rules (\ref{eqn:x})--(\ref{eqn:v}).
The joint pdf reads
\begin{multline}
  P(\sigma^\I,\sigma^\II|\beta,\lambda)=\delta(\sigma^\I-\mathrm{sign}(s_0+\tilde{M}_0^\I))\cdot \\ \delta(\sigma^\II-\mathrm{sign}(-s_0+\tilde{M}_0^\II))\cdot \rho_G(\tilde{M}_0^\I-\tilde{M}_0^\II)\;, \label{eqn:pjoint} 
\end{multline}
so that the pdf (\ref{eqn:rhosigma}) takes the Bell's form (\ref{eqn:bellint})
with $P(\lambda|\beta)=1/2$ and thus MI satisfied.
However, the joint pdf (\ref{eqn:pjoint}) is clearly not factorizable (FC fails) due to the third factor that depends on $\mu_0$ and both on $\beta^\I$ and $\beta^\II$.
In fact, while OI assumption is satisfied since $P(\sigma^\I|\beta,\lambda)$ does not depend on $\sigma^\II$ (and vice versa), PI assumption is not fulfilled since $P(\sigma^\I|\beta,\lambda)$ depends explicitly on both $\beta$'s through the coincidence condition.

With an alternative viewpoint, one may take directly $s^\RR$ as the outputs. In this case, FC would be satisfied, but one should recognize that the relevant HV are now defined for a subset of the probability space $\Lambda$,
\begin{equation}
    \Lambda_C=\{\lambda:|T^\I(\beta^\I,\lambda)-T^\II(\beta^\II,\lambda)|<W\}\;.
\end{equation}
Therefore the conditional probability  $P(\lambda|\beta)$ at the right-hand side of (\ref{eqn:bellint}) now becomes
\begin{equation}
 \rho(\lambda|\beta^\I,\beta^\II)=\rho_G(\tilde{M}_0^\I-\tilde{M}_0^\II)\rho(s_0)\rho(\mu_0)\;,  
\end{equation}
where the probability of coincidence has been explicitly considered. In this case, the MI is clearly not satisfied since $P(\lambda|\beta)\neq P(\lambda)$.

Within both viewpoints, the proposed model is not forbidden by Bell's theorem, which is based on Bell's assumptions, to violate Bell's inequalities.
Overall, the non-validity of this assumption cannot be explained with the usually  solutions (i.e., either non-localism or non-realism), nor with an equally-unpleasant form of (super-)determinism \cite{vervoort}.

\tgreen{Local-realistic models exploiting the coincidence loophole have 
been ruled out by some recent experiments that were designed to close the loophole \cite{aguero2012,giustina2013}. These experiments (conducted with polarized photons instead of spins) use a fixed time slot (FTS) method to count coincidences, in conjunction with a coincidence loophole-free Bell-type inequality.
In doing that, they are not vulnerable to the coincidence loophole \cite{larsson2013}.}
%
%
\tgreen{Nevertheless, it is the author's opinion that providing a physically sound local realist model, that explicitly realizes the loophole in question and violates the BI, strengthens at least the argument that these loopholes need to be taken seriously.}

Note that a similar \tgreen{argument} was discussed in \cite{2017} for momentum-entangled systems, although in that case the two observables (detector positions $x^\I$ and $x^\II$) are not binary functions. Still, a crucial role is played by the definition of coincidences, in this case based on the simultaneous (expected) arrival of the two particles at localized detectors, which makes the model not able to be expressed in Bell form, i.e., with either MI or FC not satisfied.

\section{Conclusions}
The paper has shown that non-relativistic quantum mechanics including spin properties can be reproduced by realistic, stochastic, and localistic  rules applied to individual particles of an ensemble. 
QM predictions are indeed retrieved as probability distributions of position, momentum, angular momentum, spin, etc. without appealing to the QM mathematical machinery itself. 

To represent spin scenarios, such as Stern-Gerlach apparatuses or a Bell test experiment, the proposed model does not appeal to two-dimensional complex spinors and matrices but uses a relatively simple set of rules, implying that (1) spin is a dichotomic quantity carried on by particles whose value depends on a random source setting and a spin propensity; (2) the latter can vary at each iteration as a function of polarization, which is a three-dimensional attribute, and the magnetic field experienced; (3) polarization is randomly attributed during preparation and has its own rules of variation; it further concurs in determining how particles react to magnetic fields.

Not being based on Bell's assumptions, but instead exploiting what is commonly considered a loophole of that theorem, the local and realist features of the model proposed do not prevent it to violate Bell's inequalities and recover experimental correlations for entangled systems as predicted by quantum mechanics.


\appendix

\section{Summary of the Local-Realistic Model for a Spinless Particle}\label{sec:summ}

In this section, we summarize the general model rules that, in the considered particular case, lead to (\ref{eqn:x})--(\ref{eqn:v}). We describe rules for particle emissions (\ref{sec:emis}), microscopic motion (\ref{sec:mot}), and how probability densities are derived from them (\ref{sec:prob}). The reader is referred to \cite{2017} or its companion paper \cite{compa} for more detail.

\subsection{Lattice and particle emissions} \label{sec:emis}
The lattice is composed of three spatial dimensions $x=\{x_d\}\in \mathbb{Z}^3$, and one temporal dimension $n\in \mathbb{N}$. Each of the spatial dimensions is characterized by the same fundamental length and acts independently. 

We describe ensembles of particles that are emitted at some sources after having been similarly prepared. 
Source setting consists in defining the number of sources $N_s$, their location $x_0^{(k)}=\{x_{0d}\}\in \mathbb{Z}^3$, probabilities $P_0^{(k)}$ (such that $\sum_k P_0^{(k)}=1$), and phase $\epsilon_0^{(k)}=\{\epsilon_{0d}\}\in\mathbb{Q}^3$, with $k\in[1,N_s]$. 
For each particle of the ensemble a source $k$ is chosen according to their probabilities. Thus preparation fixes the initial position $x_0$ and phase $\epsilon_0$.
Additionally, the source momentum $v_0=\{v_{0d}\}\in \mathbb{Q}^3$, and the source (momentum) polarization $\rho=\{\rho_d\}\in\mathbb{Q}^3$ are randomly attributed to the particle. The latter two quantities are further subject to the conditions $\sum_{d=1}^3\rho_d^2=1$, $\sum_{d=1}^3 v_{0d}^2\leq 1$.

\subsection{Microscopic motion} \label{sec:mot}
Microscopic motion is defined by a set of rules involving quantities carried by particles and quantities carried by lattice nodes (subscript $xt$).

The particle-carried quantities are: its span $\ell=\{\ell_d\}\in\mathbb{Z}^2$, lifetime $t\in \mathbb{N}$, momentum $v=\{v_d\}\in \{-1,0,1\}^3$, momentum propensity $\bm{v}=\mathbb{E}[v]$, energy propensity $e=\mathbb{E}[v^2]$, quantum momentum $v_Q=\{v_{Qd}\}\in\mathbb{Q}^3$, momentum due to external forces $v_F=\{v_{Fd}\}\in\mathbb{Q}^3$. 

Particles exchange momentum-mediating entities called ``bosons'' with the lattice, according to the mechanism illustrated below. External-force bosons (FB) carry a momentum $f=\{f_d\}\in\mathbb{Q}^3$ \tred{and phase $\epsilon_f\in\mathbb{Q}^3$}, while quantum particle bosons (PB) carry momenta $w^{(\cdot)}\in\mathbb{Q}$ and their own lifetime $t^{(\cdot)}$.

The particle's motion rules are summarized as
\begin{eqnarray}
&& t[n]=t[n-1]+1, \quad t[n_0]=0 \label{eqn:tn}\\
&& \ell_d[n]=\ell_d[n-1]+v_d[n], \quad \ell_d[n_0] = 0 \label{eqn:ell}\\
&& x_d[n]=x_{0d}+\ell[n]
\label{eqn:xn}\\ 
&& \mathbb{P}(v_d[n]=\pm 1)=\frac{e_d[n]\pm \bm{v}_d[n]}{2}\;,\quad \mathbb{P}(v_d[n]=0)=1-e_d[n] \label{eqn:prv}\\
&& e_d[n]=\frac{1+\bm{v}_d^2[n]}{2} \label{eqn:ene}\\
&& \bm{v}_d[n]=v_{Qd}[n]+v_{Fd}[n]
\label{eqn:vprop}\\
&& v_{Fd}[n]=\sum_{n'=n_0+1}^n f_d(x[n'],n') \label{eqn:force}\\
&& v_{Qd}[n]=v_{0d}-\rho_d^2 \sum_\ell \sum_{\lambda\neq \ell} w^{(\ell\lambda)}[n] \label{eqn:vq}\\
&& \tred{\epsilon[n] = \epsilon_0+ \sum_{n'=n_0+1}^n \epsilon_f(x[n'],n')\;,}\label{eqn:ereps1}\\
&& w^{(\ell\lambda)}[n]= w^{(\ell\lambda)}[n-1] \cdot\left(1-\frac{1}{2\tQR}\right) \label{eqn:pbm}
\end{eqnarray}
where $n_0$ is the iteration at which the emission has taken place.

Equations (\ref{eqn:tn})--(\ref{eqn:xn}) describe the increments of lifetime, span, and position as a function of momentum. Equations (\ref{eqn:prv})--(\ref{eqn:ene}) relate the probability distribution of momentum to momentum propensity. Equation (\ref{eqn:vprop}) states that momentum propensity is the sum of two contributions, due to quantum and external forces, respectively. 
External forces are described by interactions with the lattice, where each node can be occupied by a force boson. When a particle visits the node, it captures the resident FB and incorporates its momentum as described in (\ref{eqn:force}). A new boson is then recreated at the node.
In (\ref{eqn:vq}), quantum momentum is initially set to the source momentum and then build up from an exchange of bosons and their momenta between the particle and the lattice. The dynamics of the PB-momentum is given in (\ref{eqn:pbm}).

Lattice-carried quantities are the span trace $\lambda_{xt}=\{\trace\}\in\mathbb{Z}^3$ and the phase trace, $\epsxt\in\mathbb{Q}^3$, which represent the memory of the span and phase carried by the last particle that has visited the node $x$ with lifetime $t$. Additionally, the exchange with particles generate quantum lattice bosons (LB), carrying momenta $\omega_{xt}^{(\cdot)}\in\mathbb{Q}$, whose dynamics read
\begin{eqnarray}
&& \omg[n]=\omg[n-1]\cdot\left(1-\left(\frac{\omg[\nQR]}{\tQR}\right)^2\right)\;. \label{eqn:lbm}
\end{eqnarray}

Rules (\ref{eqn:tn})--(\ref{eqn:lbm}) are partially overcome in case of a quantum reset or an external reset. A Quantum Reset (QR) occurs when $\ell_d \neq \trace$ for at least one dimension $d$. If it is the case, the following exchanges apply:
\begin{eqnarray}
&& w^{(\ell\lambda)} \xLeftarrow[QR]{} \frac{\omega_{xt}^{(\ell\lambda)}}{\sum_{d=1}^3\rho_d^2 \delta_d^{(\ell\lambda)}} \label{eqn:pb}\\
&& \omg \xLeftarrow[QR]{} \sum_{d=1}^3 \delta_d^{(\ell\lambda)} v_{Qd} -\epsilon^{(\ell\lambda)} \label{eqn:lb}\\
&& \ell_d \xLeftrightarrow[QR]{} \trace \label{eqn:swspan}\\
&& \epsilon \xLeftrightarrow[QR]{} \epsxt  \label{eqn:sweps}
\end{eqnarray}
where $\delta_d^{(\ell\lambda)} := |\ell_d-\trace|$ is the path difference and $\epsilon^{(\ell\lambda)} := \epsilon-\epsxt$ is the phase difference. 

Rules (\ref{eqn:pb})--(\ref{eqn:lb}) state that the QR creates a new momentum-carrying LB, labeled $\ell\lambda$ to unambiguously identify the information carried by the particle, resp., the lattice node. The new LB replaces the old one of the same type, which is transferred to the particle and becomes a particle boson (PB). Rules (\ref{eqn:swspan})--(\ref{eqn:sweps}) describe the exchange of variables between the particle and the lattice.

An External Reset (ER) occurs when an external-force boson is captured, and is defined by the rules
\begin{eqnarray}
&& v_{d0} \xLeftarrow[ER]{} \bm{v}_d \label{eqn:erv0}\\ && \ell_d \xLeftarrow[ER]{} \ell_d-2f_d \frac{\sum_{d'=1}^3 \ell_{d'}f_{d'}}{\sum_{d'=1}^3 f_{d'}^2}
\label{eqn:er} \\ && \epsilon \xLeftarrow[ER]{} \epsilon+1\;.\label{eqn:ereps}
\end{eqnarray}
Although not necessary, rule (\ref{eqn:erv0}) is introduced here for the sake of model elegance. It states that each ER can be seen as a new emission, thus removing the special role of sources that are now seen as just the nodes where the last interaction has taken place. 
Rule (\ref{eqn:er}) generalizes the 1D situation where the span's sign in inverted at each external interaction. Rule (\ref{eqn:ereps}) adds a $\pi$ phase angle after each interaction.

Note that rules (\ref{eqn:ell})--(\ref{eqn:ene}) can be summarized as (\ref{eqn:x}) with $\bm{v}$ rewritten as $v$.
Moreover, in the absence of quantum forces ($v_Q\equiv v_0$), rules (\ref{eqn:vprop})--(\ref{eqn:force}) clearly correspond to (\ref{eqn:v}).

\subsection{Probability densities} \label{sec:prob}
The source position, momentum, and polarization are treated as random variables. In particular, the probability density function of the source momentum is  $\rho(v_{0d})=(1/2)$, $\forall d$. 

Stochastic preparation implies that $\bm{v}$ and thus $x$ are random variables, too. We aim at evaluating the probability mass function of the position for an ensemble of similarly-prepared particles. Unfortunately, it is generally not possible to explicitly evaluate $\rho(x;t)$. However, as discussed in \cite{2017}, for sufficiently large times we can use the approximation $\rho(x;t)\approx \rho(\bm{x};t)$, where $\bm{x}\in\mathbb{R}^3$ is the expected value of the position\footnote{We shall generally denote expected values with bold letters.}.
We describe in the rest of this section the procedure to evaluate the joint pdf's $\rho(\bm{x};t)$ and $\rho(\bm{v}_Q)$ in the presence of quadratic potentials. 

It was shown in \cite{2017} that, in case of a quadratic potential, the expected motion is given by
\begin{equation}
    \bm{x_d}=A_d(t)x_{0d}+B_d(t)\bm{v}_{Qd}+C_d(t)\;, \label{eqn:xd}
\end{equation}
with
\begin{align}
\bm{v}_{Qd} &= v_{0d}-\rho_d\sum_i \sum_{j\neq i} \sqrt{P_0^{(i) } P_0^{(j)}} \frac{\sin(arg)}{\pi\sum_{d=1}^3 \rho_d\delta_d^{(ij)}}
\label{eqn:vqm}
\\
(arg) &= \pi\sum_{d=1}^3 \delta_d^{(ij)}  \frac{\bm{x_{d}}-A_d(t)\frac{x_{0d}^{(i)}+x_{0d}^{(j)}}{2}-C_d(t)}{B_d(t)}-\pi\epsilon^{(ij)}\;.	\label{eqn:arg}
\end{align}
where $A(t)$, $B(t)$, and $C(t)$ are functions of lifetime whose form depends on the FB momentum (external force) $f(x,t)$. For example, a free particle is described by $A=1$, $B=t$, $C=0$; a free faller by $A=1$, $B=t$, $C=ft^2/2$; an harmonic oscillator by $A=\cos\Omega t$, $B=\sin\Omega t/\Omega$, $C=0$.
Equations (\ref{eqn:vqm})--(\ref{eqn:arg}) are valid as the limit after a sufficiently long number of iterations and emissions (particle and lattice ``training" as defined in \cite{2017}).

The joint pdf of the positions is found by applying the rule
\begin{equation}
    \rho(\bm{x};t)=\frac{1}{2}\left|\frac{\partial (v_{01},\ldots,v_{03})}{\partial(\bm{x}_1,\ldots,\bm{x}_3)}\right| \;,
\end{equation}
yielding
\begin{equation}
    \rho(\bm{x};t) = \frac{1+\sum_i \sum_{j\neq i} \sqrt{P_0^{(i) } P_0^{(j) } }  \cos \left(arg \right) }{\prod_{d=1}^3 2|B_d(t)|} \label{eqn:rhox3d}
\end{equation}
In \cite{2017} it was shown that the Schr\"{o}dinger equation and Born rule can be retrieved from (\ref{eqn:rhox3d}).

Similarly, the joint pdf of the momentum propensities is evaluated as
\begin{equation}
    \rho(\bm{v}_Q;t)=\frac{1}{2^3} \left(1+\sum_i\sum_{j\neq i}\sqrt{P_0^{(i)} P_0^{(j)}}  \cos\left( \pi\sum_{d=1}^3 ⁡\delta_d^{(ij)}\bm{v}_{Qd}-\pi \epsilon^{(ij)}\right) \right) \;. \label{eqn:rhov}
\end{equation}

In (\ref{eqn:x})--(\ref{eqn:v}), the momentum propensity was considered as a constant for all particles in the ensemble. The implicit assumption there was that the particles were prepared according to a ``wavepacket" preparation (and that both lattice and particle training are achieved). 

Wavepackets can be prepared by setting a finite number $N_{s}$ of sources at adjacent nodes centered at $x=0$. The source phase is set as $\epsilon_d(x)= mx$, with $m\in[0,1]$. The probability is set as constant, $P_{0}(x)=1/N_{s}$ for a ``plane wave" preparation or as $P_{0}(x)=\displaystyle\frac{1}{2^{N_{s}-1}}{N_{s}-1 \choose x+\frac{N_{s}-1}{2}}$
for a ``Gaussian wave" preparation with variance $(N_{s}-1)/4$.

When (\ref{eqn:rhov}) is applied to these source probabilities and phases, one obtains a momentum probability density that peaks for $\hat{v}_{Q}=m$. The peak intensity increases with $N_{s}$ for plane waves, with $\sqrt{N_{s}}$ for Gaussian waves, as it is the case with QM states. 
For a sufficiently large $N_s$ this pdf can be practically treated as a Dirac delta function, whence the implicit assumption behind (\ref{eqn:x})--(\ref{eqn:v}) with $m$ there denoted as $v_0$.
%

\section{Higher spins} \label{sec:higher}

In this section we extend the model of Sect.~\ref{sec:spin} to particle having spin $S$ higher than \textonehalf. For particles with general spin number $S$ ($S=1/2, 1, 3/2, \ldots$), the quantity here denoted as spin ($s$) can take $2S+1$ values equispaced between +1 and -1 ($s\in[1,1-1/S,\ldots,-1]$).

\subsection{Microscopic motion}

In addition to their source spin and $\mu$-polarization, particles are emitted with a second polarization vector (third, including $\rho$) $\tau_0=\{\tau_{0d}\}\in\mathbb{Q}^3$.
The constraint on polarization components is generalized as 
\begin{equation}
  \sum_{d}\mu_{d0}^2 = \tilde{s}_0^2, \quad \sum_{d}\tau_{d0}^2 = \frac{S+1}{S}-\tilde{s}_0^2 \;,
  \label{eqn:mu0squared}
\end{equation}
where $\tilde{s}_0$ denotes the spin-rounded value of $s_0$, $\mathrm{Round}(S(1+s_0))/S-1$.
Clearly, for $S=1/2$, one retrieves  $\tilde{s}_0=\pm 1$ and $\sum_d\mu_{d0}^2=1$.

To describe the dynamics of polarization, equation (\ref{eqn:polarev}) still holds when applied to both vectors $\mu$ and $\tau$. In the presence of a constant magnetic field, these dynamics conserve the quantities $\sum_d\mu_d^2$ and $\sum_d\tau_d^2$.

The rules replacing (\ref{eqn:s_def}) determine at each iteration the particle's spin $s$ among the $2S+1$ possible values. 
The general rule reads
\begin{equation}
    s[n] = -1+\frac{1}{S}\sum_{k=1}^{2S}\Theta(s_0-\tilde{s}_k[n]), \quad \tilde{s}_k:=-1+2\sum_{\sigma=-1}^{-1+\frac{k-1}{S}} P(\sigma)
\end{equation}
where $\Theta$ is the unit step function and $P(\sigma)=\mathbb{P}(s=\sigma)$ is the spin pmf. 
Clearly, (\ref{eqn:s_def}) is retrieved for $S=1/2$, with $P(\pm 1)=(1\pm M)/2$ .

The pmf of spin (a $2S+1$-valued discrete random variable) is completely defined by its first $2S$ central moments.
The latter are prescribed by the model as functions of the quantities $M=\sum_d \mu_d\beta_d$ and $T:=\sum_d \tau_d\beta_d$.
The first two central moments read
\begin{eqnarray}
    && \mathbb{E}[s] = M \;, \label{eqn:esm} \\
    && \mathbb{E}\left[(s-M)^2\right] := V = \frac{\sum_d\tau_d^2-T^2}{2} \;, \label{eqn:var}
\end{eqnarray}
with $\mathbb{E}\left[(s-M)^3\right] = -\mathbb{E}[(s-M)^2]\cdot \mathbb{E}[s]$ etc. 
Overall, we can write $s[n]=f(M[n],V[n],s_0)$.

The ER rule (\ref{eqn:er}) is still valid. Additionally,
\begin{equation}
\tau_d \xLeftarrow [\mathrm{ER}]{}     \beta_d\sqrt{\frac{S+1}{S}-s^2}
    \label{eqn:tau}
\end{equation}
holds. 
Note that, since $\sum_d\beta_d^2=1$ by definition, both $\sum_d\mu_d^2$ and $\sum_d\tau_d^2$ are constant across an ER.

\subsection{Probability densities}

For a homogeneous and constant magnetic field, (\ref{eqn:rho_up}) is generalized as follows. Due to the polarization dynamics, the quantities $M$ and $T$ do not change with time, thus $M[n]\equiv M_0$, $T[n]\equiv T_0$. Similarly, $\sum_d\mu_d^2$ and $\sum_d\tau_d^2$ are constant. The transition probabilities are thus constant, too. Since $s_0$ is attributed to a particle once and for all, the spin $s$ will stay constant while crossing the magnetic field $\beta$. Its probability distribution is thus uniquely determined by the values $M_0$ and $T_0$.

For a non-uniform magnetic field (SG apparatus), again the probability distribution of spin is determined at the first ER, that is, at the entry of the SG, and is a function of the quantities $M_0$ and $T_0$. With the ER (\ref{eqn:tau}), $M$ jumps to $s$ and $T$ jumps to $\sqrt{\frac{S+1}{S}-s^2}$. Consequently, the spin propensity $\mathbb{E}[s]$ is $s$ and the variance becomes zero. At the next iteration, the spin will remain constant with probability one. The probability of spins at the exit of the SG is thus uniquely determined by the values $M_0$ and $T_0$.

In the case of two SG in cascade, we have $\mu_1=s_1\beta_1$ and $\tau_1=\beta_1\sqrt{\frac{S+1}{S}-s_1^2}$ at the exit of the first SG. Therefore, at the entry of the second SG, $M_1=s_1Y$ and $V_1=\left(\frac{S+1}{S}-s_1^2\right)\frac{1-Y^2}{2}$, having defined $Y:=\sum_d\beta_{1d}\beta_{2d}=\cos(\beta_1,\beta_2)$.
Accordingly, the spin takes a value $s_2=f(M_1,V_1,s_0)$. After the first ER, the spin propensity jumps to $s_2$ and the variance jumps to zero. Therefore, at the next iterations, the spin will remain equal to $s_2$. Overall, the probability of having a certain spin $s_2$ at the exit of the second SG depends on $Y$ and $s_1$.

As an example, for $S=1$, the spin pmf is evaluated from the first two moments as
\begin{equation}
    P(1)=\frac{V+M^2+M}{2}, \quad 
    P(0) = 1-V-M^2, \quad P(-1)=\frac{V+M^2-M}{2} 
    \label{eqn:s1}
\end{equation}
Thus the spin pdf at the exit of the second SG is
\begin{equation}
 \rho(s_2)=(1-s_2^2)+\frac{s_2}{2}M_1+\left(-1+\frac{3}{2}s_2^2\right)(V_1+M_1^2) \;.
\end{equation}
Using the expressions for $M_1$ and $V_1$ derived above, one obtains
\begin{equation}
    \rho(s_2|s_1)=\left(\frac{s_2^2-s_1^2}{2}-\frac{3}{4}s_1^2s_2^2\right)+\frac{s_1s_2}{2}Y+\left(1-\frac{3}{2}(s_1^2+s_2^2) +\frac{9}{4}s_1^2s_2^2 \right)Y^2 \; \label{eqn:rhoS1}
\end{equation}
which precisely match the QM results computed, e.g., in \cite{tekin}\footnote{Cascaded Stern-Gerlach probabilities for higher spin have been seldom studied in the literature. Equations (\ref{eqn:esm})--(\ref{eqn:var}) and, consequently, equations of the type (\ref{eqn:rhoS1}), are not present in \cite{tekin} but have been built upon the general result of that paper.}. Further, it can be verified that $\rho(s_2|s_1)=\rho(s_1|s_2)$, as it should be.

\end{document}